# Emergent exotic chirality dependent dielectricity in magnetic twisted bilayer system

Yu-Hao Shen[1], Wen-Yi Tong[1], He Hu[1], Jun-Ding Zheng[1], Chun-Gang Duan[1,2*]

[1] *State Key Laboratory of Precision Spectroscopy and Key Laboratory of Polar Materials and Devices, Ministry of Education, Department of Electronics, East China Normal University, Shanghai, 200241, China*

[2] *Collaborative Innovation Center of Extreme Optics, Shanxi University, Taiyuan, Shanxi 030006, China*

E-mail: cgduan@clpm.ecnu.edu.cn

## Abstract

Twisted van der Waals bilayers provide an ideal platform to study the electron correlation in solids. Of particular interest is the 30° twisted bilayer honeycomb lattice system, which possesses an incommensurate Moiré pattern and uncommon electronic behaviors may appear due to the absence of phase coherence. Such system is extremely sensitive to further twist and many intriguing phenomena will occur. In this work, we show that due to the twist induced spatial inhomogeneity of interlayer coupling, there emerges an $U(1)$ gauge field in magnetic transition-metal dichalcogenides (TMD) bilayers. Interestingly, for further twist near 30°, the induced gauge field could form a chirality dependent real-space skyrmion pattern, or magnetic charge. Moreover, such twist also induces the topology dependent electronic polarization of the bilayer system through the nonzero flux of the real-space Berry curvature. Further analysis proves that the antiferromagnetically coupled twisted bilayer system is indeed also antiferroelectric! When an external electric field is applied to break the potential balance between layers, there will emerge novel magnetoelectric coupling and exotic chirality dependent dielectricity. Such findings not only enrich our understanding on Moiré systems, but also open an appealing route toward functional 2D materials design for electronic, optical and even energy storage devices.

**Introduction**

Twistronics rooted in the twisted bilayer van der Waals crystals is of both theoretical and technological importance[1-6]. Recent advances in fabrication of atomically thin materials have successfully realized the interlayer twist, giving rise to an alternative way to modulate layered potential. In such twisted 2D systems, Moiré pattern with long period[7-13] is introduced by misoriented stacking. The patterned interlayer coupling in van der Waals crystals significantly modifies the low-energy band structure. For instance, it is characterized by the flat bands in twisted bilayer graphene (tBLG) and the Dirac velocity become zero several times with the change of the magic-angle[10,14-16]. The strong electron correlation in tBLG yields various fascinating physical behaviors, such as the transitions from semimetal to Mott insulator and even unconventional superconductivity[14,16-18].

Unlike the magic-angle tBLG system with commensurate superlattice, when the twist angle is 30° between the two layers, the singularity appears, making the system lack of long period [19-22] and become incommensurate with 12-fold rotational symmetry. More interestingly, for the bilayer transition-metal dichalcogenides (TMD) 2H-MX$_2$ (M=Mo, W and X=S, Se) system [23-31], due to the existence of different stackings, turning left and right near 30° twist angle corresponds to different atomic structures. Consequently, the two generated Moiré systems possess different phase modulations of the electronic wavefunction in the superlattices, rendering chirality dependent properties that are absent in tBLG system[21,23,24]. For instance, the modulation on exchange interaction[32,33] and the Berry curvature[34-36] form different patterns in two twisted cases.

Further, as magnetism is extremely sensitive to the structural change in two-dimensional systems[33,37,38], we expect the non-spin-degenerate twisted bilayer system with large intrinsic exchange coupling could demonstrate more fascinate electron correlation behaviors. Such considerations therefore forge us to extend the Moiré TMD bilayers to magnetic system.

In this work, based on a typical magnetic system 2H-VSe$_2$, we show that the low energy Hamiltonian of such system of the twisted bilayer case satisfies $SU(2) \times U(1)$ symmetry. Here, the twist induced emergent $U(1)$ gauge field is chirality dependent, forming apparently different skyrmion patterns for the case of left and right twist near 30°. Due to the connection between the Berry curvature and the electronic polarization we show that the twist will induce opposite polarization in top and bottom layers, i.e. the system is indeed antiferroelectric! When we apply a vertical electric field to break the potential balance between layers, there will emerge exotic chirality dependent dielectricity and novel magnetoelectric coupling. Further investigations demonstrate that the deep origin of the special dielectric response is from the topology protection of the pseudospin structure in these two Moiré systems. All above theoretical results are proved numerically by *ab-initio* calculations. The potential applications using such dielectric properties are proposed as well.

**Results and discussion**

We choose the ferromagnetic 2H-VSe$_2$, a typical 2D ferrovalley material[39,40], as a platform and set its bilayer to be antiferromagnetically coupled, which is the ground state of most pristine bilayer system[41,42]. Taking 30±2.2° twisted cases for example and the in-plane shift between the two layers is set to be zero. The top view of the incommensurate (infinite) and commensurate (finite) lattice structures are shown in Fig.1a, in which the supercell of the two commensurate partners possess the same size (Fig. 1b). Here we define that the δ<0 (>0) case corresponds to the right (left) rotation of top layer with respect to bottom layer in the 30+δ° twisted bilayers. Note that the band structures for left and right twisted bilayer systems correspond to different superpositions of the monolayer bands. This can be understood by the mirror symmetry of the two first Brillouin Zone (BZ) of the supercell along $\mathbf{\Gamma} - \mathbf{K}$ axis[24], illustrated as the grey hexagons in Fig. 1b. In considering of the bands folding into this first BZ, we obtain that for left twisted supercell, $\mathbf{K}_+^s$ states is superposed by $\mathbf{K}_+$ valley states of bottom layer and $\mathbf{K}_-$ valley states of top layer (H-type like stacking), whereas for the right twisted supercell, it is superposed by the two $\mathbf{K}_+$ valley states of both two layers (R-type like stacking).

As the interlayer hopping of the valley electron depends sensitively on the atomic configuration between the layers, in our case there emerges a gauge field dependent on the smooth variation of the interlayer registry to describe the electron-electron interaction of its low energy effective Hamiltonian[34-36]. Starting from the previous work[43-45] on tBLG which has $SU(4)$ gauge symmetry, we additional introduce the valley and spin quantum numbers $v$ and $\sigma$. The low energy exchange Hamiltonian of the magnetic TMD bilayers can be written as:

$$H_{ex} = \sum_{i \neq j} \sum_{\sigma,\sigma'} J_{ij} ( c_{iv\sigma}^+ c_{jv\sigma'}^+ c_{iv\sigma'}^- c_{jv\sigma}^- + c_{iv'\sigma}^+ c_{jv'\sigma'}^+ c_{iv'\sigma'}^- c_{jv'\sigma}^- + c_{iv\sigma}^+ c_{jv'\sigma'}^+ c_{iv'\sigma'}^- c_{jv\sigma}^- ) \quad (1)$$

Here, $J_{ij} > 0$ is the exchange coupling, $c_i^{+(-)}$ denotes the creation (annihilation) operator for an electron and satisfy the anti-commutation relation $\{c_i^+, c_j^-\} = 2\delta_{ij}$. Eq. (1) can be further cast into layer dependent Hamiltonian with $SU(2) \times U(1)$ gauge symmetry (see Sec. I of SI for details):

$$H_{ex}^\ell = -2 \sum_{i \neq j} J_{ij} \left[ \frac{1}{2} + \boldsymbol{\tau}_i \cdot \boldsymbol{\tau}_j + \tau_i^z \tau_j^z + i\ell(\boldsymbol{\tau}_i \times \boldsymbol{\tau}_j) \cdot \boldsymbol{e}^z \right] \left[ \frac{1}{4} + \boldsymbol{S}_i \cdot \boldsymbol{S}_j \right] \quad (2)$$

Here $\ell$ is the layer index. The cross-product term is chirality and layer dependent due to the spin-valley locking[46-49] and spin-layer locking effect[37,50]. We choose $\ell = -1$ and $1$ to represent top and bottom layer, respectively. $\boldsymbol{S}$ is the localized spin operator and $\boldsymbol{\tau}$ is the valley pseudospin operator. We notice that $\tau_i^z \tau_j^z + i\ell(\boldsymbol{\tau}_i \times \boldsymbol{\tau}_j) \cdot \boldsymbol{e}^z$ is a complex number operator, which can be regarded as a $U(1)$ gauge field $e^{i\psi}$ after normalization. Here, the $\psi$ describes the canting of the spatial inhomogeneity induced pseudospin ($\boldsymbol{M}$) of one layer in the superlattice. Apparently, the Hamiltonian (Eq. (2)) is $U(1)$ gauge invariant, i.e. it has $U(1)$ symmetry.

To exactly describe the pseudospin $M$ induced by spatial distribution of a twist, we choose the specific gauge for our case with triangular symmetry[34,51]. It is $\psi = \boldsymbol{G}_i^\parallel \cdot \boldsymbol{r} + \frac{\pi}{2}$ ($i = 1,2,3$), where the $\boldsymbol{G}_i^\parallel$ is the reciprocal lattice vector of the superlattice obtained by counterclockwise rotation of $\boldsymbol{G}_1^\parallel = \frac{4\pi}{\sqrt{3}a_M} \boldsymbol{e}^y$ with angle $(i-1)\frac{2\pi}{3}$ ($a_M$ is the lattice constant of the Moiré systems and $\boldsymbol{e}^y$ is the unit vector of $y$ axis). Then for 30+δ° twisted bilayers, $M$ is dependent on the position $r$ as:

$$\boldsymbol{M}_z = M_0 \sum_{i=1,2,3} \cos(\boldsymbol{G}_i^\parallel \cdot \boldsymbol{r} + \frac{\pi}{2}) \boldsymbol{e}^z \tag{3a}$$

$$\boldsymbol{M}_\parallel = M_0 \sum_{i=1,2,3} \sin(\boldsymbol{G}_i^\parallel \cdot \boldsymbol{r} + \frac{\pi}{2}) \boldsymbol{e}_i^\parallel \tag{3b}$$

Here, $\boldsymbol{e}_i^\parallel$ is the corresponding unit vector of $\boldsymbol{G}_i^\parallel$. $M_0$ characterizes the amplitude of $M$.

Note that the $\boldsymbol{G}_i^\parallel$ of left and right twisted cases are orthogonal with each other in the same coordinate reference.[24] For simplicity we define $\boldsymbol{r} = (x, y)$ and $\boldsymbol{r}' = (y, -x)$ for left and right twisted system, respectively. The distribution of the $M$ are shown in Fig. 2a. Surprisingly, the in-plane pseudospins form totally different vortex structures (skyrmion lattice) in left and right twist systems. For the left twisted case, it corresponds to the Néel skyrmion lattice texture. Whereas it becomes the Bloch one for the right twisted case, whose helicity phase γ has $\frac{\pi}{2}$ difference[52]. In addition, the analytically calculated topological winding number[52] $N_{sk} = \frac{1}{4\pi} \iint d^2r \frac{\boldsymbol{M}}{|\boldsymbol{M}|^3} \cdot (\partial_x \boldsymbol{M} \times \partial_y \boldsymbol{M})$ are identical (= 1) for both cases. As we expected, the vortex pattern of $M$ agrees with that of the pseudospin texture $M_k$ obtained by our *ab-initio* calculations (Fig. S3). Such conclusion is independent of the choice of the gauge, or $\psi$. The real space Berry curvature $\Omega_r$ of the Moiré superlattice[35] can be obtained from $M$. Specifically, the skyrmion lattice of $M$ specifically forms an emergent pseudomagnetic field as[52]:

$$\Omega_r^z = \frac{1}{2} \frac{\boldsymbol{M}}{|\boldsymbol{M}|^3} \cdot (\partial_x \boldsymbol{M} \times \partial_y \boldsymbol{M}) \tag{4}$$

The distribution of $\Omega_r^z$ is shown in insets of Fig. 2a, whose maximum is centered at the origin. Since $\Omega_r$ covers a unit sphere, it can be equivalently regarded as a pseudomagnetic field generated by a magnetic charge, or magnetic monopole ($= \frac{1}{2}$).

Considering the relationship between the Berry curvature and the electronic polarization, we find that the in-plane inhomogeneous pseudomagnetic field will induce an out-of-plane electronic polarization $\Delta \boldsymbol{P}_e$, which can be written as[53,54]:

$$\Delta P_e = \frac{e}{(2\pi)^2} \int_0^1 [\iint \varepsilon_{\lambda k_x k_y} (A_\lambda \partial_{k_x} A_{k_y} - i\frac{2}{3} A_\lambda A_{k_x} A_{k_y}) dk_x dk_y] d\lambda \qquad (5)$$

where we assume the adiabatic transformation (twist by small $\delta$) is parametrized by a scalar $\lambda$ and $\varepsilon_{\lambda k_x k_y}$ is a totally antisymmetric tensor. In our case, only the highest occupied band has significant Berry curvature. The Berry connection $A_\eta = i\langle u_{k\lambda} | \partial_\eta u_{k\lambda} \rangle$ ($\eta = \lambda, k_x, k_y$). Here, $|u_{k\lambda}\rangle$ is the eigen state of the pseudospin $\boldsymbol{M}$ as $\lambda$ evolves from 0 to 1. As mentioned above, in left and right twisted cases, $\boldsymbol{M}$ forms a skyrmion with same winding number (= 1)[52]. Thus, the magnetic flux $\oint \boldsymbol{\Omega}_k \cdot d\boldsymbol{S}_k = \oint \boldsymbol{\Omega}_r \cdot d\boldsymbol{S}_r$ is independent on $\lambda$ and always equals to $2\pi$ for both cases. Further analysis (see Sec. III of SI) shows that the electronic polarization $\Delta P_e = \alpha \delta$ with $\alpha$ being a coefficient. Above discussion is based on one spin-channel. Due to the spin-valley and spin-layer locking effect in the present bilayer system, the other spin-channel also has electronic polarization $\Delta P_e(\delta) = -\alpha \delta$. Consequently, the antiferromagnetically coupled twisted bilayer system should be antiferroelectric, rendering the system to be multiferroic! In addition, from the calculated charge distribution difference along $z$ axis between the two twisted cases $\Delta \rho = \rho_R - \rho_L$ (Fig. 2b), we can determine the direction of layer-resolved polarization for different twisted systems (as shown in Fig. 3c).

It is known that, the bilayer system is sensitive to the interlayer electric bias as it breaks the perpendicular inversion symmetry [55-57] and the energy degeneracy of two layers[42,58]. To describe this effect, we then develop a two-band $\boldsymbol{k} \cdot \boldsymbol{p}$ model for the low energy valence states of the 30+δ° twisted bilayer, which is identical for spin up and down states for such antiferromagnetically coupled bilayer systems under external electric field:

$$H(\boldsymbol{k}) = \begin{pmatrix} -\frac{\hbar^2 k^2}{2m^*} - \frac{V_0}{2} - \frac{V_z}{2} & I \\ I & -\frac{\hbar^2 k^2}{2m^*} + \frac{V_0}{2} + \frac{V_z}{2} \end{pmatrix} \qquad (6)$$

Here $\boldsymbol{k}$ is momentum measured from the $\boldsymbol{K}$ valley and $m^*$ is the valence band effective mass under parabolic approximation. $I$ is the intralayer magnetic exchange parameter and it appears in the off-diagonal term due to the antiferromagnetic interlayer coupling. $V_0$ is created by the internal electronic polarization $\Delta \boldsymbol{P}_e(\delta)$ due to twist. $V_z$ is created by the external electric filed $\boldsymbol{E}$ and the reference is set to be in the middle of the bilayer system. The positive $\boldsymbol{E}$ is defined from top layer pointing to bottom layer. When the electric field $\boldsymbol{E}$ reverses, the shift of the spin-up (↑) sub-bands in bottom layer would be same as the spin-down (↓) sub-bands in top layer under original $\boldsymbol{E}$, indicating that $\boldsymbol{E}$ can select the specific spin sub-bands close to the Fermi level. It can be regarded as a kind of spin-layer locking[37,50] for such antiferromagnetically coupled bilayers. For the valance bands of spin up states at $\boldsymbol{K}$ valley, the diagonalization gives rise to the energy gap $\Delta$ between upper and lower band, which is from different layers (see the Sec. II of SI for detailed derivation):

$$\Delta = 2I + \frac{1}{4I}(V_z + V_0)^2 \qquad (7)$$

Here, $V_0$ is negative and positive for left and right twisted case, respectively. In our case, the change of $\Delta_\uparrow$ and $\Delta_\downarrow$ corresponds to the case of positive and negative $E$, repectively. The change of $\Delta$ with $V_z$ is therefore an even function. From Eq. (7), we can see clearly that the right and left twist cases possess amplified and suppressed dielectric response, respectively due to the coupling between twist induced internal potential $V_0$ and external electric potential $V_z$.

Moreover, as is well know, the topology of the skyrmion lattice is robust. The pseudospin structures of the twisted bilayers are thus expected to be topologically protected under external perturbation. Therefore, under small $E$, the modification of the Berry curvature $\mathbf{\Omega}$ is merely through the shift of the energy level of the system, and the topological connectivity of the pseudospin wavefunction remains unchanged. Based on the Kubo formula[53,59,60] and using the relationship between the Berry phase and the electronic polarization[61,62], we can obtain the analytical relationship between $E$ and its induced electronic polarization $\Delta P(E)$ as (see Sec. IV of SI):

$$\Delta P(E) = P(E) - P(0) = \frac{e\mathbf{c}}{V}\left(1 - \frac{\Delta^2(0)}{\Delta^2(E)}\right) = \frac{e\mathbf{c}}{V\Delta^2(0)}(2V_0 V_z + O(V_z^2)) \tag{8}$$

Here, $\mathbf{c}$ is the lattice vector along the $z$ axis and $V$ is the supercell volume. When $V_z$ is small, we obtain the right twisted bilayer exhibits positive polarization $\Delta P > 0$ ($V_0 > 0$), whereas the left twisted bilayer exhibits negative polarization $\Delta P < 0$ ($V_0 < 0$), which is stunning since this means that the susceptibility is *negative*!

To prove the above predictions, we then carry out *ab-initio* calculations to study the effect of vertical electric field on the band structure of the two systems (Fig. S3). The so obatained band shift is exactly consistent with the analytical prediction using the Eq. (7) (see Fig. 3a), demonstrating the validity of the two-band $\mathbf{k}\cdot\mathbf{p}$ effective Hamiltonian we adopted to describe the electronic properties of the twisted bilayer under external electric field. Moreover, the calculated electronic polarization $\Delta P(E)$ by *ab-initio* method, shown in the inset of Fig. 3a, also exactly agrees with the analytical prediction of Eq. (8) (the inset of Fig .3a). Taking $E = 0.001$ V/Å as an example, the calculated results are $\Delta P = -0.98$ and $\Delta P = 1.26$ for left and right twisted case, respectively. Here, we define the $P$ as the dipoles/volume per layer, which is in units of $10^{-4}$ $e$/Å$^2$. Compared with the calculated value $\Delta P = 1.03$ of the monolayer system, the twisted bilayers obviously possess suppressed or amplified dielectric polarization. The abnormal dielectric response of the bilayer systems is more clearly seen from the distortion of the electron clouds, as illustrated in the charge difference plot ($\Delta \rho = \rho(E) - \rho(0)$, shown in Fig. 3b). For left twisted case there induces negative polarization, i.e. the induced dipole is, as predicted by the model analysis, *opposite* to the applied electric field, in strong contrast to that of the right twisted case, where there is almost no induced dipole that exceeds the monolayer thickness. Indeed, this kind of response is also very strange. As we know, due to the electrostatic effect, the induced dipoles in general are always along the direction of the applied field, for not only the dielectrics but also the magnetic metal film[63,64].

The above unusual phenomenon can be explained clearly by the different dielectric response from the Moiré systems with different skyrmion pattern of the pseudospin under external electric field. Note that the effective two-band Hamiltonian of the Bloch skyrmion lattice (right twist, δ<0) and Néel skyrmion lattice (left twist, δ>0) are rather different (shown in Table 1). The Bloch one has a Rashba-like term $(\boldsymbol{\sigma} \times \boldsymbol{k}) \cdot \boldsymbol{e}_z$. When $\boldsymbol{E}$ is turned on, the electron tends to leave the layer due to the electrostatic force. However, it would decrease the planar electron density and weaken the intralayer magnetic exchange interaction (Δ), which is unfavorable to decrease the total energy. To avoid this situation, the planar electron density rearranges to strengthen the magnetic exchange interaction by increasing (decreasing) the curling of the valley pseodospin ($\boldsymbol{\sigma} \times \boldsymbol{k}$) in the bottom (top) layer locked with spin up (down) electrons. The net positive polarization is then constructed between layers (right panel of Fig. 3c). Therefore, through the enhancement of planar magnetic exchange interaction, the system can reduce the energy cost by the introduction of the external electric field. And the perpendicular change of electron density, especially the part out of the layer, is negligible. That is why there is almost no induced dipole exceeding the monolayer thickness in this case (right panel of Fig. 3b). In addition, like the electric field effect, the increase of the magnetic exchange coupling further separates the bands of top and bottom layers. As a result, the bands split under electric field is doubled, which can be regarded as electric field amplifying effect.

Whereas for the left twist (δ>0) with Néel skyrmion lattice, the effective two-band Hamiltonian has no Rashba-like term. As a result, there is no way to change the curling of the pseodospin to respond the external electric field, as it will change the topological helicity phase (γ). The only choice left to protect the topological valley pseodospin structure as well as to keep the total energy minimum is to remain Δ unchanged (Table 1). Consequently, the in-plane electrons move collectively out-of-plane to screen or resist the external electric field and therefore *negative* polarization is formed, as clearly shown in the yellow arrows of left panel of Fig. 3c. Such mechanism keeps the system as close as possible to the unperturbed state ($\boldsymbol{E} = 0$). And we can see that the *negative* susceptibility, or the electric field resisting effect, is indeed deeply related to the topologically protected skyrmion lattice with certain $N_{sk}$ and γ. The negative electric susceptibility for left twisted case is unusual and often appears in metamaterials using artificial design. We assume this novel property may result in useful application in optical waveguide or other optical and electronic devices, e.g., negative capacitor[65,66]. We find that in either twisted case, the electrostatic energy will be converted in to magnetostatic energy. Therefore, such systems have quite large energy capacity.

Finally, we should point out that though in this study we take 2H-VSe$_2$ bilayer as example to demonstrate the twist effect, it should be ubiquitous for any magnetic bilayers with stacking dependent electronic structures. The magnetic twisted system responds to the external electric field in a magnetic way, which can be regarded as an entirely new magnetoelectric effect, reveals us an intricate pattern how electrostatic and magnetostatic interactions can be entangled when the system become complicated.

**Conclusion**

We show that twist can induce an emergent gauge field in the magnetic bilayer systems. It is demonstrated that this gauge field forms a chirality dependent real-space skyrmion pattern, or equivalent magnetic charge, for further twist near 30°. Moreover, the nonzero magnetic flux of the real space Berry curvature will generate a pure electronic polarization, whose sign is opposite for antiferromagnetically coupled top and bottom layers, i.e., such twisted magnetic bilayer system is indeed also antiferroelectric. When an external electric field is applied, there will occur exotic chirality dependent dielectricity for the left and right twisted cases, dramatically different from that of monolayer. This unique phenomenon well interprets the philosophy of *More is different*[67]. This kind of novel magnetoelectric effect brings unusual phenomena to the system, e.g. *negative* susceptibility, direct electrostatic-magnetostatic energy transforming. We propose several potential applications based on the exotic dielectric behaviors and expect more inspiring discoveries on these twisted systems.

**Method**

When the $\delta = \pm 2.2°$, the corresponding supercell possesses the $\sqrt{13} \times \sqrt{13}$ times size of the unit cell[68], as shown in Fig S1. For the monolayer 2H-VSe$_2$, the intermediate layer of hexagonally arranged vanadium atoms are sandwiched between two atomic layers of selenium. The optimized lattice constant is 3.335Å. From top view, it possesses honeycomb structure and is lack of inversion symmetry. When turned from $\delta = 2.2°$ case to $\delta = -2.2°$ case, for the bilayer structure, the top layer stacked on the bottom layer rotates 180° along the [110] of its supercell. Then we obtain two different stacking bilayers, where they possess twist angle 32.2° and 27.8°.

The *ab-initio* calculations of bilayer VSe$_2$ are performed within density functional theory (DFT) using the projector augmented wave (PAW) method implemented in the Vienna *ab-initio* Simulation Package (VASP)[69]. The exchange-correlation potential is treated in Perdew–Burke–Ernzerhof form[70] of the generalized gradient approximation (GGA) with a kinetic-energy cutoff of 400 eV. A well-converged 9×9×1 Monkhorst-Pack k-point mesh is chosen in self-consistent calculations. The convergence criterion for the electronic energy is $10^{-5}$ eV and the structures are relaxed until the Hellmann–Feynman forces on each atom are less than 1 meV/Å. In our calculations, the spin-orbit coupling (SOC) effect was explicitly included in the calculations and the dispersion corrected DFT-D2 method[71] is adopted to describe the van der Waals interactions. The external electric field is introduced by planar dipole layer method.

**Acknowledgements**

This work was supported by the National Key Research and Development Program of China (2017YFA0303403), Shanghai Science and Technology Innovation Action Plan (No. 19JC1416700), the NSF of China (No. 51572085, 11774092), ECNU Multifunctional Platform for Innovation.

**Figures**

**Figure 1** | (a) Top view of the bilayer 2H-VSe$_2$ commensurate case of left twist (left panel), right twist (right panel) and incommensurate case of 30° twist case (middle panel). Grass-green and red sites of the honeycomb represent bottom layer selenium and vanadium atoms, respectively. Yellow and blue sites in the hexagons represent top layer selenium and vanadium atoms, respectively. (b) The first Brillouin Zone (BZ) of the superlattice is shown as the grey hexagons.

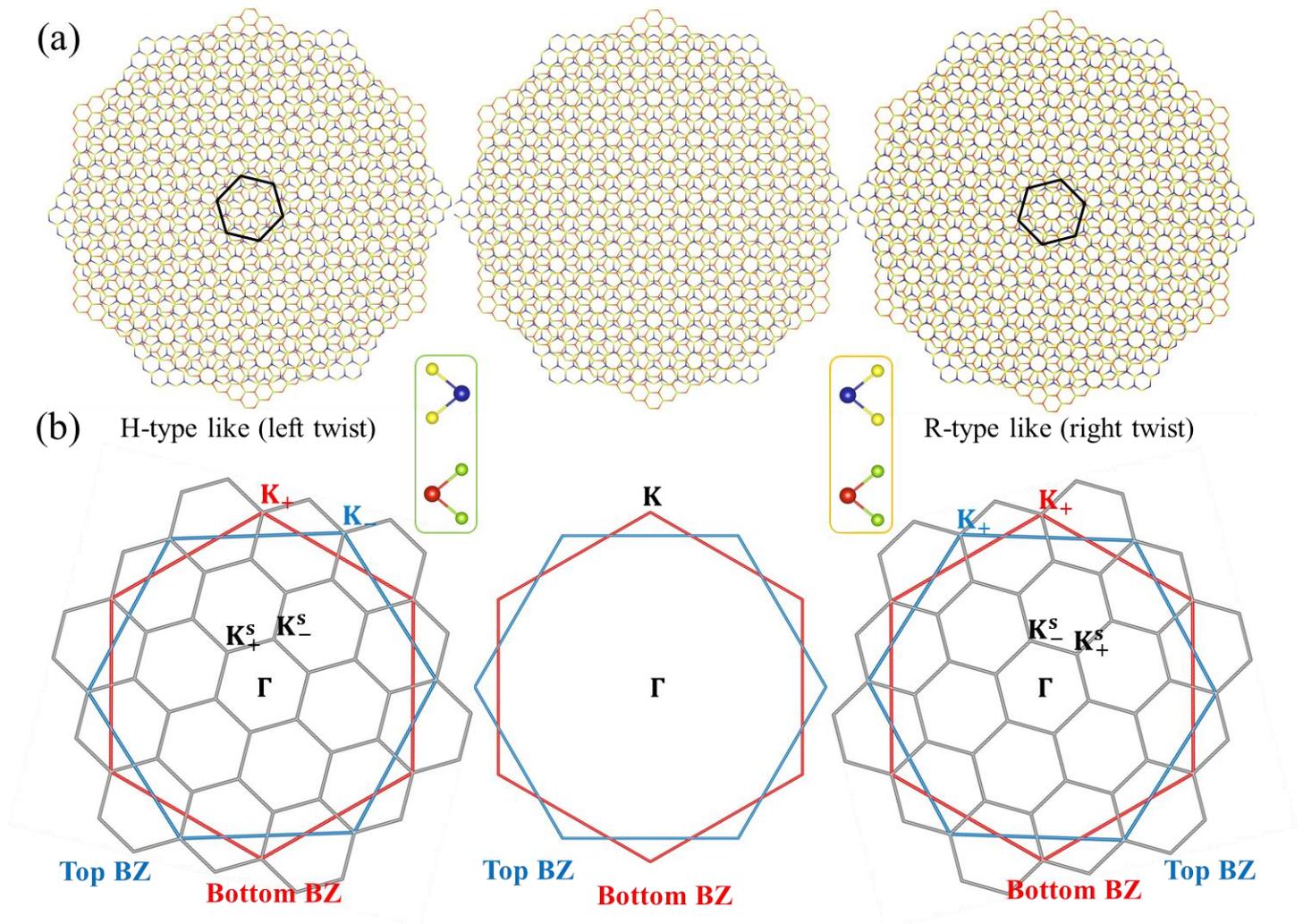

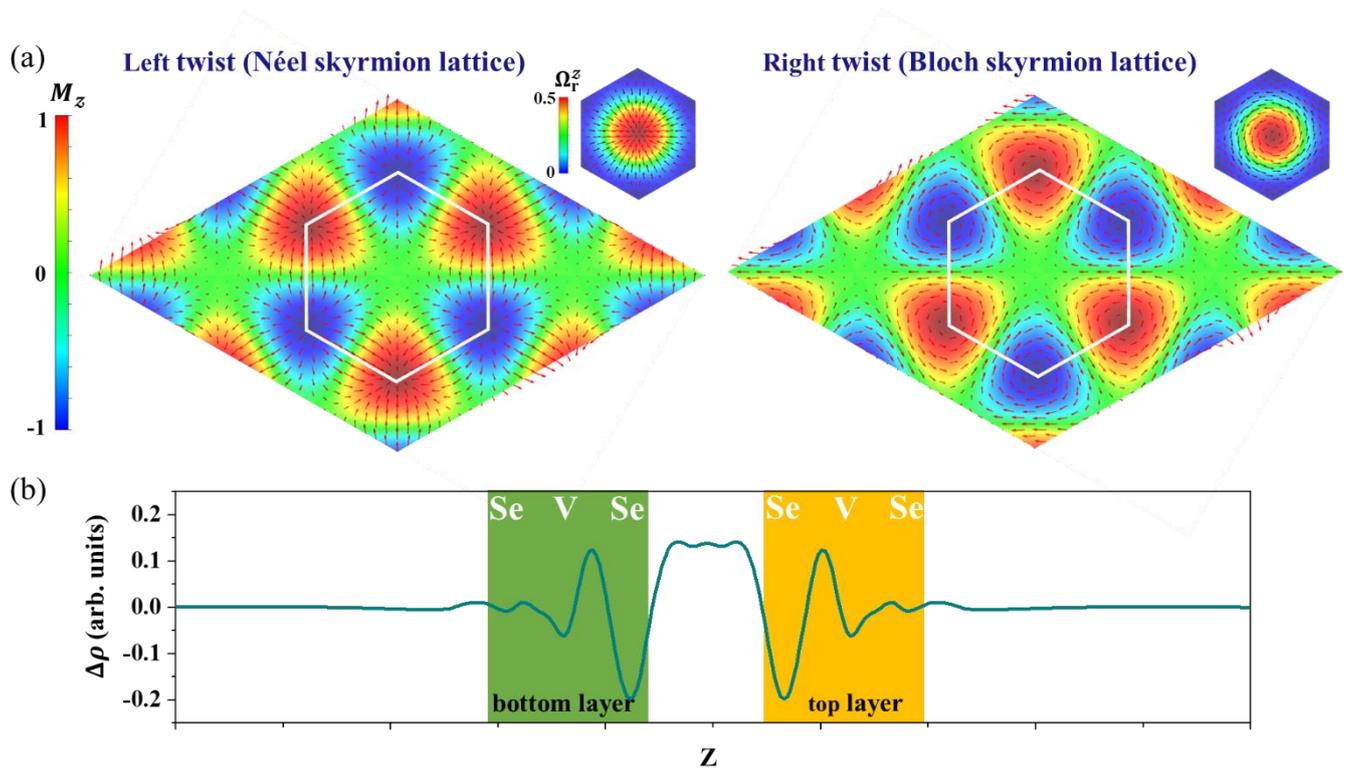

**Figure 2** | (a) Normalized layer pseudospin skyrmion pattern in the real space. The inset figures show the distribution of $\Omega_r$ in supercell denoted by the white hexagon. The in-plane and out-of-plane components are shown in red arrows and color map, respectively. (b) Induced average electron charge, $\Delta\rho = \rho_R - \rho_L$ along $z$ axis when the bilayer system turned from right (R) to left (L) twist.

**Figure 3** | (a) The evolution of Δ as a function of electric field $E$ by fitting the *ab-initio* calculated results denoted by circles. We choose the values of parameters in Eq. (7) as $I = 0.456$ eV and $V_0 = 0.113$ eV. The inset figures show the calculated $\Delta P(E)$. (b)The induced planar charge of the bottom layer under electric field $E = 0.001$ V/Å for left twist case (left) and right twisted case (right). The electric field-induced charge densities $\Delta\rho = \rho(E) - \rho(0)$ in arbitrary units for the two twisted cases. Yellow and cyan colors represent the accumulation and depletion of electrons, respectively. Orange arrows denote the directions of the electric field. (c) The schematic electrical effect on the twist induced internal dipole (blue arrows).

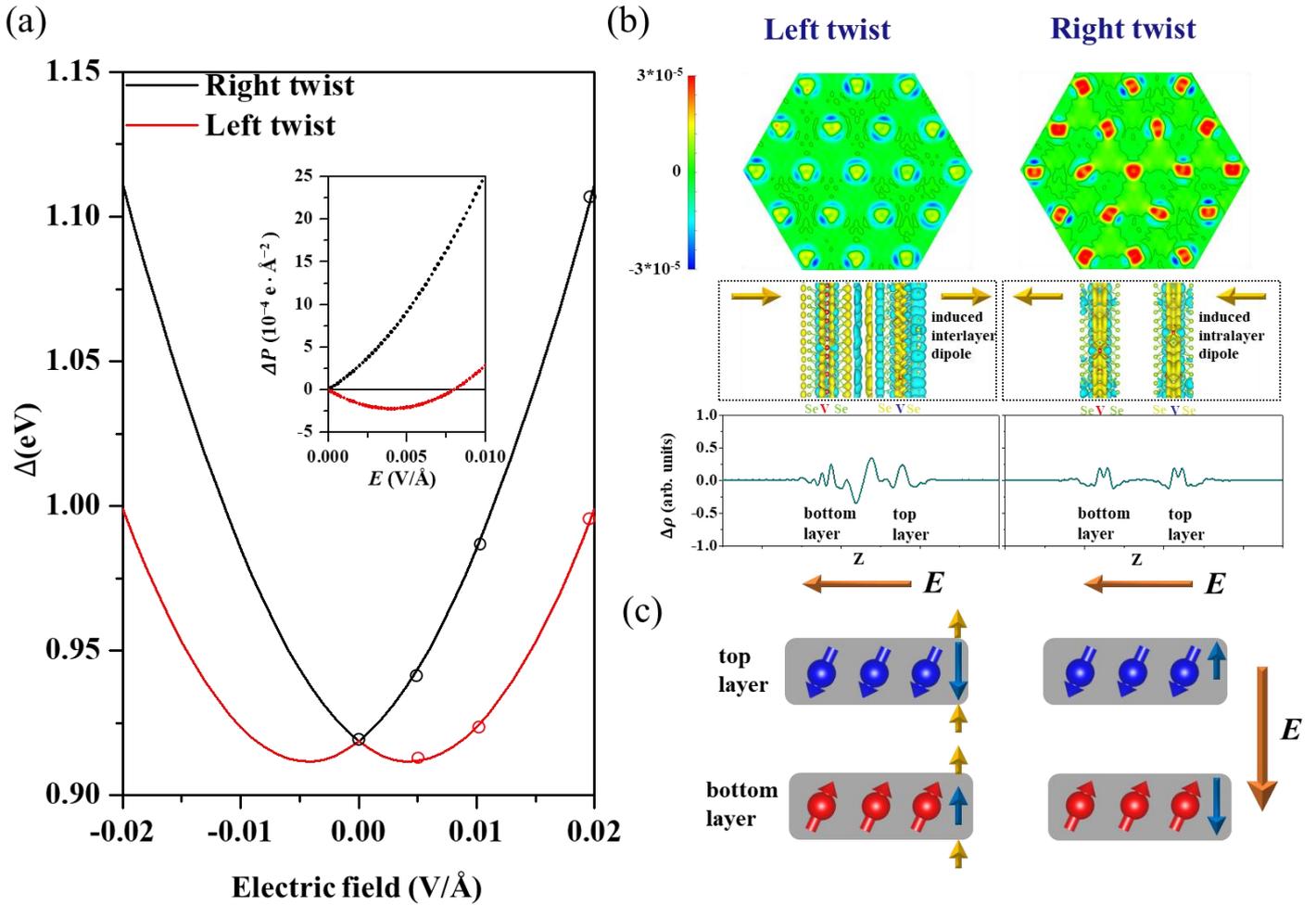

**Table 1** The two-level effective Hamiltonian of the pseudospin skyrmion lattice when $E = 0$ and its corresponding eigen states. In our case, the $\boldsymbol{k}$ space covering a unit sphere can be written as $\boldsymbol{k} = (\sin\theta\cos\varphi, \sin\theta\sin\varphi, \cos\theta)$ with introduction of the polar angle $\theta$ and azimuthal angle $\varphi$.

|  | Bloch Skyrmion | Néel Skyrmion |
|---|---|---|
| **Two-band effective Hamiltonian** | $H_0 = -\frac{1}{2}\Delta(0)[(\boldsymbol{\sigma} \times \boldsymbol{k}) \cdot \boldsymbol{e}_z + k_z\sigma_z]$ | $H_0 = -\frac{1}{2}\Delta(0)(\boldsymbol{\sigma} \cdot \boldsymbol{k})$ |
| **Eigen states** | $\|u_{\boldsymbol{k}}^+\rangle = \begin{pmatrix} \cos\frac{\theta}{2} \\ e^{i(\varphi+\frac{\pi}{2})}\sin\frac{\theta}{2} \end{pmatrix}, \|u_{\boldsymbol{k}}^-\rangle = \begin{pmatrix} \sin\frac{\theta}{2} \\ -e^{i(\varphi+\frac{\pi}{2})}\cos\frac{\theta}{2} \end{pmatrix}$ | $\|u_{\boldsymbol{k}}^+\rangle = \begin{pmatrix} \cos\frac{\theta}{2} \\ e^{i\varphi}\sin\frac{\theta}{2} \end{pmatrix}, \|u_{\boldsymbol{k}}^-\rangle = \begin{pmatrix} \sin\frac{\theta}{2} \\ -e^{i\varphi}\cos\frac{\theta}{2} \end{pmatrix}$ |